\documentclass{elsart}

\usepackage{natbib}

\usepackage{graphics}

\usepackage{amssymb}

\begin{document}

\begin{frontmatter}


\title{SOLAR FLARES WITH AND WITHOUT CORONAL MASS EJECTIONS AND TYPE II SHOCKS}
\thanks[Thanks2]{Deceased.}
\author{A. Hillaris\corauthref{UOA}},
\corauth[UOA]{A. Hillaris} \ead{ahilaris@cc.uoa.gr}
\author[TEILAM]{V. Petousis}, 
\author[UOA]{E. Mitsakou}, 
\author[UOA]{C. Vassiliou}, 
\author[UOA]{X Moussas}, 
\author[UOA]{J. Polygiannakis\thanksref{Thanks2}},
\author[UOA]{P. Preka-Papadema}, 
\author[DI]{C.Caroubalos},
\author[UOI]{C.Alissandrakis},
\author[TEILAM]{P.Tsitsipis},
\author[TEILAM]{A.Kontogeorgos},
\author[PARIS]{J.-L.Bougeret} 
\and
\author[PARIS]{G.Dumas} 

\address[UOA]{Section of Astrophysics, Astronomy and Mechanics, Department of Physics, University of
Athens, GR-15784 , Panepistimiopolis Zografos, Athens, Greece}
\address[TEILAM]{Department of Electronics, Technological Education Institute of Lamia, Lamia, Greece}
\address[DI]{Department of Informatics, University of Athens, GR-15783 Athens, Greece}
\address[UOI]{Section of Astro-Geophysics, Department of Physics, University of Ioannina, GR-45110 Ioannina,Greece}
\address[PARIS]{Observatoire de Paris, Departement de Recherche Spatiale, CNRS UA 264, F-92195 Meudon Cedex, France}

\begin{abstract}
An analysis is presented of a set of {\it radio rich} solar SXR flares, i.e. accompanied by
type IV or II metric radio bursts, associated with Coronal Mass Ejections and MHD
shocks, recorded as type II events, in the period 1998-2000. The relative size, impulsiveness
and energetics of these events are investigated. We find that, on the average,
the flares with type II bursts and CME's are larger, more impulsive and energetic from
the flares with type II but without CME's. The latter are more energetic than flares associated
with relatively {\it slow CME's} accompanied by type IV continua but not type
II shocks. Although a simple classification may not be readily determined the results
imply that associated energetic events originate, more often than not, from sources
with characteristics fairly well correlated.
\end{abstract}

\end{frontmatter}
\section{INTRODUCTION}\label{Intro}
The association of Coronal Mass Ejections, flares and metric radio bursts of the type II and IV family,
remains an oper research topic. The general trend is that the association propability raises in 
proportion to flare strength and CME velocity \citep{Kahler84}, yet exceptions abound:
{\begin{itemize}
\item {40\% of the class M flares are not associated with CMEs \citep{And03} while 86\% of the CMEs are
associated with class C flares (\cite{Har95}).}
\item {30\% of the coronal shocks, corresponding to metric type II bursts, are not associated 
with CMEs (\cite{Kahler84}, \cite{Classen}).}
\item {The non-type II associated CMEs are equally divided 
in {\it fast ($V_{CME}$~$>$~455 Km/sec)} and {\it slow ($V_{CME}$~$<$~455 Km/sec)} \citep{Sheeley}.}
\item {Type II bursts are fairly well associated both with intense and weak HXR bursts \citep{Pearson}.}
\end{itemize}}
In this report, based on a data set of {\it radio rich}
events, we analyse the size, impulsiveness and energetics of the associated GOES SXR recordings 
in order to explore the nature of flares apparently associated with Coronal Mass Ejections and type II bursts.
\section{OBSERVATIONAL RESULTS \& ANALYSIS} \label{Obs}
We use a medium size data set of {\it radio rich}
events, ie, CMEs and flares accompanied by metric type II or IV bursts (obtained by the radiospectrograph
ARTEMIS-IV), in the 1998-2000 period. 
The gross spectral characteristics of these events and the 
associated CME and flare parameters are summarised 
in \citet{Car2004}.\\
\begin{table*} \label{T1}
\begin{center}
\textbf {Table 1: Summary of Peak \& Integrated Flux for SXR Flares.}
\begin{tabular}{|l|lll|lll|}
\hline 
		   & Peak Flux     & (W/$m^2$)     & 		  & Integrated   & Flux         & (J/$m^2$)\\
		   & Mean          & Min           & Max          & Mean         & Min           & Max\\
\hline 
Class A 	   & 7.2~$10^{-5}$ & 4.5~$10^{-7}$ & 5.6~$10^{-4}$&8.8~$10^{-2}$ & 2.6~$10^{-4}$ & 0.75 \\
Class B 	   & 2.6~$10^{-5}$ & 2.1~$10^{-6}$ & 1.8~$10^{-4}$&1.8~$10^{-2}$ & 2.1~$0.1^{-2}$& 0.11  \\
Class C	   & 4.4~$10^{-6}$ & 7.8~$10^{-7}$ & 8.0~$10^{-6}$&0.7~$10^{-2}$ & 0.3~$10^{-2}$& 0.01  \\
\hline 
\end{tabular}
\end{center}
\end{table*}
From the forty events of the original data set data set we have eliminated all events coinciding with LASCO DATA GAPS and those without an SXR flare tabulated in the Solar Geophysical Reports. 
\begin{table*} \label{T2}
\begin{center}
\textbf {Table 2: Duration, Rise Time \& Decay Time of the SXR Flares.}
\begin{tabular}{|l|lll|lll|lll|}
\hline
		     & Rise  & Time  & (sec)  &  Decay        & Time          & (sec)	& Dur.	&(sec)	&\\
		     & Mean  & Min   & Max    & Mean          & Min           & Max	& Mean	&Min		&Max\\
\hline
Class A	     & 867.0 & 240.0 & 3480.0 & 1304.0        & 240.0         & 4980.0 	& 2177.0	& 480	 & 5710\\
Class B	     & 877.6 & 241.0 & 2340.0 &  864.0        & 216.0         & 2090.0 	& 1741.6	& 481	 & 4430\\
Class C	     &1488.0 & 540.0 & 2280.0 & 1928.4        & 540.0         & 3540.0 	& 3416.4	& 1080	& 5760\\
\hline
\end{tabular}
\end{center}
\end{table*}
The remaining thirty three events were partitioned into, twenty ejective flares associated with a type II metric burst
({\em class A\/} throughout the text), eight non-ejective flares associated with a type II metric 
burst ({\em class B\/}) and five ejective flares without a type II metric burst ({\em class C\/}).
For each class of event we analyse SXR flare data such as:
{\begin{itemize}
\item {Colour Temperature and Emission measure from the GOES measurements, following \cite{G94}}
\item {Impulsiveness, defined as the quotient of Peak Flux (1-8 A GOES channel) over the rise time and 
representing the average growth rate of the flare (\cite{Pearson}, albeit for HXR bursts, 
also \cite{Magda01}, \cite{Vrsnak}). }
\item {SXR flare characteristics: Peak flux, total integrated flux (1-8 A GOES channel) and rise time, 
which were taken from the Solar Geophysical Data. The decay time was estimated from the GOES (0.5-4 A channel)
data as the interval required for the drop of flux to 25\% of the peak value. The event duration was set equal to 
the sum of rise and decay times.}
\end{itemize}}
\begin{table*} \label{T3}
\begin{center}
\textbf {Table 3: Impulsiveness (W~$m^{-2}$~$sec^{-1}$) of the SXR Flares.}
\begin{tabular}{|l|llll|}
\hline
	   & Mean & Min & Max  & \\
\hline
Class A	 & 8.0~$10^{-8}$ & 1.9~$10^{-9}$ & 4.4~$10^{-7}$  & \\
Class B	 & 4.0~$10^{-8}$ & 2.1~$10^{-9}$ & 2.8~$10^{-7}$  & \\
Class C	 & 4.5~$10^{-9}$ & 5.7~$10^{-10}$ & 1.5~$10^{-8}$  & \\
\hline 
\end{tabular}
\end{center}
\end{table*}
The results of this analysis, pertaining to the SXR flare characteristics, are summarised in Tables 1 to 4. 
We, furthermore, examine the variations of the flare parameters as well as the CME and Type II kinetics vs 
SXR integrated flux. In figure \ref{FluxVII},
we present velocities of CMEs versus the total integrated flux (1-8 A GOES channel) of
the associated SXR flare; drift velocities of type II bursts versus the total integrated flux;
flare rise time versus decay time and emission measure as a function of colour temperature.
Certain observational results arise as follows:
\begin{itemize}
\item{The ejective flares associated with type II shocks (Class A) are more energetic 
than non ejective flares with type II bursts (Class B); these, in turn, are more energetic than ejective flares 
which are not accompanied by type II events (Class C). 
Both the average peak flux and the average time integrated SXR flux
differ almost by an order of magnitude from a category to the next, yet their ranges overlap (cf. Table 1).}
\item{The rise and decay time do not differ significantly on the average, while their ranges overlap. 
The class C events present a higher rise and decay time that the other two 
categories while duration of Class B events is smaller than the rest (cf. Table 2, also \citet{Kahler84}
for similar results). The length of the rise phase increases with the length of the decay phase in almost the 
same way for all the three event categories (cf. figure \ref{FluxVII}, bottom left, 
also \citet{Kay03}). }
\item{Ejective flares associated with type II shocks (Class A events) overcome on average impulsiveness
non ejective flares with type II bursts (Class B) by a factor 2; these, in turn, overcome the 
non-type II ejective flares (Class C) by an order of magnitude. The ranges overlap, yet the third category
appears significantly less impulsive than the first two (cf. Table 3).}
\item{The three categories, do not differ signifucantly on the average emission measure and temperature (cf Table). 
The class B and class C events have the same average temperature,
(10~$10^6$K) and are clustered in the lower emission measure-temperature range  
(cf. figure \ref{FluxVII}, bottom right). The majority of class A events on the other hand 
reaches higher temperatures and emission measures than the rest. The colour temperature-emission
measure relation is consistent with previous results by \citet{Kay03}.}
\item{The type II drift velocities appear all in excess of 400 Km/sec 
and uncorrelated with the SXR time integrated flux (cf. figure \ref{FluxVII},top left). There is not 
any apparent separation between non-eruptive flare associated type II bursts (Class B events, or flare blast
originating MHD shocks) and the CME driven shocks (Class A events, or piston driven shocks).
The scatter in the data points is attributed mostly to the dependence of the type II velocity, mainly, on 
ambient coronal conditions (cf. for example \cite{Pearson}, also \cite{Kahler84}).}
\item{The CME velocities appear fairly well correlated to the SXR Integrated flux, 
(cf figure \ref{FluxVII},  top right) The non-type II CMEs (Class C)
are clustered towards the lower velocity and flux range, yet connected to the rest
of the sample. This result is in accordance with the CME velocity versus peak flux correlation reported by 
\citet{Moon}; the position of class C events on the velocity-integrated flux plane corroborates \citet{Sheeley}
who state that type II associated CME velocities exceed a threshold of about 400 Km/sec.}
\end{itemize}
{\begin{figure*}[ht]
\resizebox{\hsize}{!}{\includegraphics{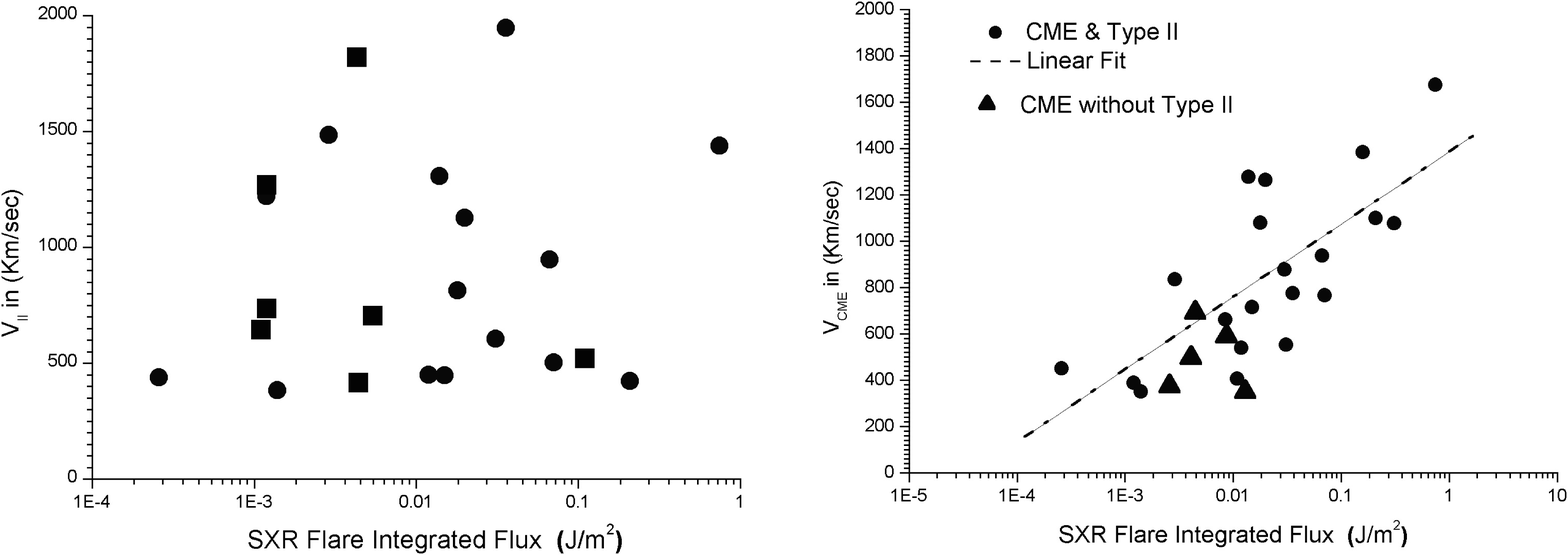}}
\resizebox{\hsize}{!}{\includegraphics{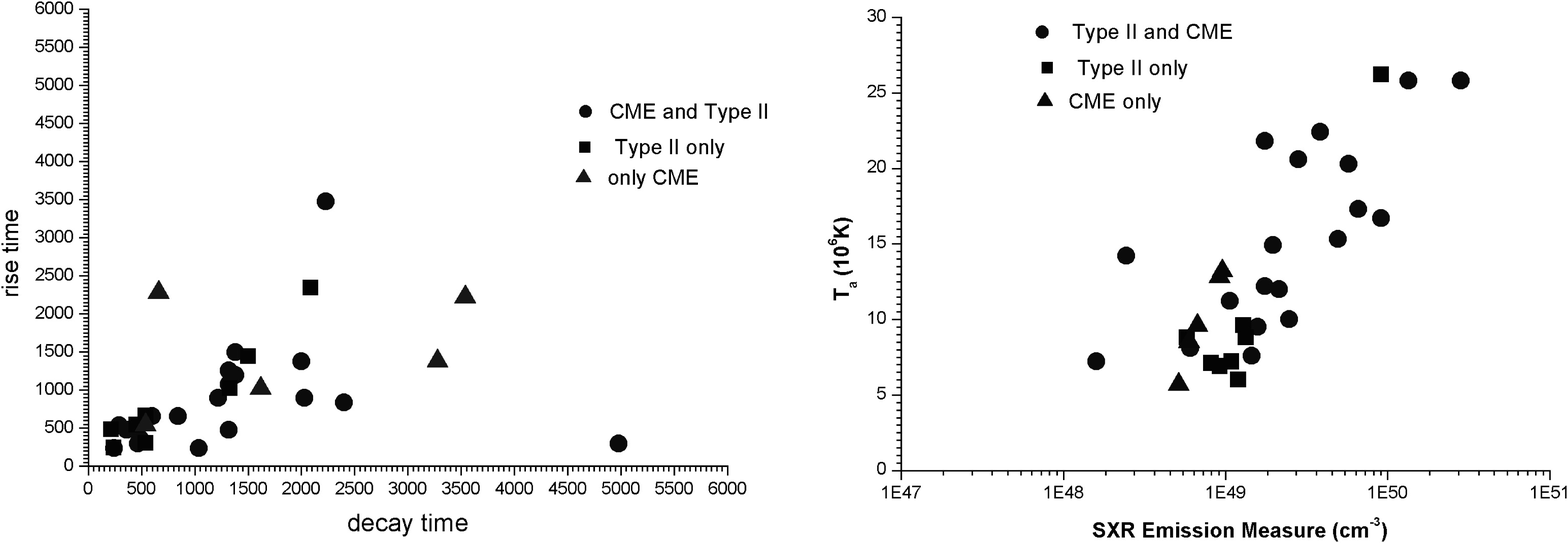}}
\caption{Top left: Type II drift velocities versus integrated SXR flux.
Top Right: CME Velocities versus integrated SXR flux. Bottom Left: SXR Rise Time vs Decay Time.
Bottom Right: Colour Temperature versus Emission Measure. Circles correspond to Flares with CMEs associated
with type II bursts (Class A events), triangles to flares with CMEs not accompanied by type II (Class C), 
and squares to flares with type II bursts not coincident with CMEs (Class B).}
\label{FluxVII}
\end{figure*}}
\begin{table*} \label{T4}
\begin{center}
\textbf {Table 4: Colour Temperature \& Emission Measure.}
\begin{tabular}{|l|lll|lll|}
\hline
               & Colour& Temp.	  & ($10^6$K)  & Emission  	 & Measure 		& ($cm^{-3}$)\\
		   & Mean  & Min  	  & Max  	   & Mean      	 & Min     		& Max\\
\hline
Class A	   & 15.1  & 7.2  	  & 25.8 	   & 4.6~$10^{49}$ & 1.6~$10^{48}$ 	& 2.8~$10^{50}$  \\
Class B	   & 10.1  & 6.0  	  & 26.2 	   & 2.0~$10^{49}$ & 5.8~$10^{49}$ 	& 8.6~$10^{49}$   \\
Class C	   & 10.0  & 5.7  	  & 13.2 	   & 7.3~$10^{49}$ & 5.2~$10^{48}$ 	& 4.4~$10^{48}$  \\
\hline
\end{tabular}
\end{center}
\end{table*}
\section{DISCUSSION AND CONCLUSIONS}\label{AR}
The comparison between the soft X-ray characteristics, has underlined a certain ordering where Class A events are
more energetic than Class B which in turn are more energetic than class C. Further more the Class B events (Type II
without CME) have the shortest duration while Class C events (Non-Type II CMEs) are the least impulsive. The ranges of
parameters, however, overlap significantly; the least energetic and the flares of low temperature are almost 
equally divided among the three classes introduced in this report, while the most energetic events are associated 
with type II shocks and CMEs. \\
The lack of a clear dividing line among classes implies that both CMEs and type IIs are present in a wide range of
SXR flare parameters. The need for a close examination of the less energetic events, almost suggests 
itself in the hope of a resolution of this uncertainty.
\ack{This work was financially supported by the Research Committee of the University of Athens.}

\end{document}